\documentclass[a4paper, fleqn, usenatbib]{mnras}
\usepackage{amssymb,amsmath}
\usepackage{graphicx}
\usepackage{xcolor}
\usepackage{multirow}
\usepackage[T1]{fontenc}
\usepackage{ae,aecompl}

\def\oiii{[O~{\sc iii}]\ }
\def\nii{[N~{\sc ii}]\ }
\def\sii{[S~{\sc ii}]\ }

\def\oi{[O~{\sc i}]\ }

\title[DBH model]
{To test dual supermassive black hole model for broad line AGN with 
double-peaked narrow \oiii lines}
\author[Zhang \& Feng]
       {Xue-Guang Zhang\thanks{Corresponding author Email: zhangxg23@sysu.edu.cn} \& 
        Long-Long Feng\\ 
       Institute of Astronomy and Space Science, Sun Yat-Sen University, 
          No. 135, Xingang Xi Road, Guangzhou, 510275, P. R. China}

\date{}

\begin{document}
\pagerange{\pageref{firstpage}--\pageref{lastpage}} \pubyear{2015}
\maketitle
\label{firstpage}

\begin{abstract}
   In this manuscript, we proposed an interesting method to test 
the dual supermassive black hole model for AGN with double-peaked 
narrow \oiii lines (double-peaked narrow emitters), through their 
broad optical Balmer line properties. Under the dual 
supermassive black hole model for double-peaked narrow emitters, 
we could expect statistically smaller virial black hole masses 
estimated by observed broad Balmer line properties than 
true black hole masses (total masses of central two black holes). 
Then, we compare the virial black hole masses between a sample of 37 
double-peaked narrow emitters with broad Balmer lines 
and samples of SDSS selected normal broad line AGN with single-peaked 
\oiii lines. However, we can find clearly statistically larger 
calculated virial black hole masses for the 37 broad line AGN with 
double-peaked \oiii lines than for samples of normal broad line 
AGN. Therefore, we give our conclusion that the dual supermassive 
black hole model is probably not statistically preferred to the 
double-peaked narrow emitters, and more efforts should be necessary 
to carefully find candidates for dual supermassive black holes 
by observed double-peaked narrow emission lines.
\end{abstract}


\section{Introduction}

     System of dual supermassive black holes is an inevitable 
stage of co-evolution of supermassive black hole and host galaxy 
\citep*{sr98, h06, h07, kh13, lr14}. With separations about kilo-pcs 
of central dual supermassive black holes, models based on central dual 
supermassive black holes can be efficiently applied to well explain 
observed double-peaked narrow emission lines. Therefore, the observed 
double-peaked narrow emission line can be commonly used as an indicator 
for central dual black holes, such as the reported dual supermassive 
black hole candidates based on observed double-peaked narrow emission 
lines combining with properties of high quality images 
\citep*{zw04, gn07, cg09, xk09, rs10, cp11, fm11, bs12, bce13, cs13, 
lc13, wc14}. However, there are many other studies \citep*{lg10, 
fc11, sl11, cg12, fy12} which have shown that scenarios 
with a single AGN can also well explain observed double-peaked narrow 
emission lines which could be due to radial flows or due to disk 
structures of central narrow emission line regions. 

   So far, it is controversial on the origin of double-peaked narrow 
emission lines, due to system of central dual supermassive black holes 
(hereafter, DBH model) or due to gas dynamic structures of narrow 
emission line regions in a single AGN (hereafter, GD model). However, 
one distinct point can be found between the two proposed models. The 
DBH model has apparent and strong effects on central broad emission 
lines from two independent broad emission line regions rotating 
around central dual black holes, besides apparent effects on narrow 
emission lines. However, the GD model has NO effects on central broad 
emission lines. Therefore, to check properties of broad emission lines 
of double-peaked narrow emitters will provide further information 
on the origin of double-peaked narrow emission lines. 

   There are so far more than 3000 low-redshift double-peaked narrow 
emitters reported in the literature. The main four large samples and 
corresponding detailed discussions on properties of double-peaked 
narrow emission lines can be found in \citet{wc09} (87 
double-peaked narrow emitters), in \citet{ss10} (148 
double-peaked narrow emitters), in \citet{gh12} (about 3030 
double-peaked narrow emitters) and in \citet{bl13} (131 
double-peaked narrow emitters). However, there are so far no studies 
on broad emission lines of double-peaked narrow emitters in the 
literature. Here, we first report interesting studies on properties 
of broad Balmer lines of a sample of 37 double-peaked 
narrow emitters selected from SDSS (Sloan Digital Sky Survey) 
QSOs, in order to test the DBH model for double-peaked narrow 
emission lines. This paper is organized as follows. Section 2 
shows our main hypotheses based on the DBH model, and Section 3 
gives our main results for the 37 double-peaked narrow emitters 
with broad Balmer emission lines selected from SDSS and further 
discussions on their properties of virial black hole masses, 
and then Section 4 gives our main conclusions. And in this 
manuscript, cosmological parameters 
$H_{0}=70{\rm km\cdot s}^{-1}{\rm Mpc}^{-1}$, $\Omega_{\Lambda}=0.7$ 
and $\Omega_{m}=0.3$ have been adopted.  

\section{Hypotheses}

    Based on the DBH model for double-peaked narrow emission lines, 
each observed broad emission line of a double-peaked narrow emitter 
actually includes two components from central two independent BLRs 
(broad emission line regions) with separation about kilo-pcs. The 
separation of central dual black holes is large enough, so that for 
each black hole plus one BLR system, its central virial black hole 
mass can be well determined by broad line width and continuum 
luminosity under the widely applied virialization assumption 
\citep*{md04, pf04, gh05, vp06, kb07, vo09, rh11, sr11, hk15, ww15},
\begin{equation}
\begin{split}
M_{{\rm 1}} &= k_{{\rm BH}}\times (V_{{\rm 1}})^2\times(\lambda L_{{\rm 1}})^{\sim0.5}\\
M_{{\rm 2}} &= k_{{\rm BH}}\times (V_{{\rm 2}})^2\times(\lambda L_{{\rm 2}})^{\sim0.5}\\
\end{split}
\end{equation},
where $k_{{\rm BH}}$ is a scale factor, $V$ means broad 
line width (second moment $\sigma$ or full width at half 
maximum FWHM), $\lambda L$ represents continuum luminosity at 
5100\AA\ which can be used to estimate distance of BLR to 
central black hole (the well-known R-L relation, \citet*{kas00, 
wz03, kas05, dp10, bd13}). Therefore, central total black hole 
mass of a double-peaked narrow emitter is 
$M_{{\rm tot}} = M_{{\rm 1}} + M_{{\rm 2}}$. In addition, in 
observed spectra of double-peaked narrow emitters, broad emission 
lines can not be clearly divided into two broad components, because 
their peak shifted velocities are less than several hundreds kilometers 
per second, more smaller than broad line widths. Here, due to more 
larger separations of central dual black holes in double-peaked narrow 
emitters, we accept that there are the same peak shifted velocities 
of the two broad components as those of the double-peaked narrow 
emission lines. Therefore, similar as Equation (1), central black 
hole mass of a double-peaked narrow emitter can also be estimated 
through line parameters of observed broad emission lines,
\begin{equation}
M_{{\rm BH}} = k_{{\rm BH}}\times (V_{{\rm obs}})^2\times(\lambda L_{{\rm obs}})^{0.5}
\end{equation}.
Then, properties of $M_{{\rm tot}}$ and $M_{{\rm BH}}$ should provide 
further information on the origin of double-peaked narrow emission lines.

   It is clear that for a double-peaked narrow emitter, the first 
direct point under the DBH model we can have is  
\begin{equation}
\lambda L_{{\rm obs}} \sim \lambda L_{{\rm 1}} + \lambda L_{{\rm 2}}
\end{equation},
where $\lambda L_{{\rm obs}}$ means the observed continuum luminosity at 
5100\AA. And then, because broad line width is more larger than 
peak separation of the two expected broad components under the DBH 
model, the second point we can have is 
\begin{equation}
V_{{\rm obs}}^2\sim f_{{\rm 1}}\times V_{{\rm 1}}^2 + f_{{\rm 2}}\times V_{{\rm 2}}^2
\end{equation}
where $V_{{\rm obs}}$ means broad line width measured from observed 
broad emission lines, and $f_{{\rm 1}}$ and $f_{{\rm 2}}=1-f_{{\rm 1}}$ 
mean flux ratios of the two expected broad components to total 
broad emission line ($f_{{\rm 1}}$ and $f_{{\rm 2}}$ are 
values less than 1). Here, we should note whether second moment 
or FWHM is used as broad line width ($V$, $V_{{\rm obs}}$), the 
equation above can be well accepted, unless there are much large 
peak separations. And the equation above can be directly obtained 
from definition of second moment \citep{pf04}, similar as 
what we have done in \citet{zh11}. 

   Based on the two points above, we can find that virial black 
hole mass from observed broad line parameters can be described as 
\begin{equation}
\begin{split}
M_{{\rm BH}}^2 & = f_{{\rm 1}}^2\times V_{{\rm 1}}^4\times\lambda L_{{\rm 1}} + f_{{\rm 2}}^2\times V_{{\rm 2}}^4\times\lambda L_{{\rm 2}} \\
         & = f_{{\rm 1}}^2\times M_{{\rm 1}}^2 + f_{{\rm 2}}^2\times M_{{\rm 2}}^2 
\end{split}
\end{equation}.
Therefore, a direct and interesting result is that the estimated 
virial black hole mass $M_{{\rm BH}}$ from observed broad emission 
lines is more smaller than the true total black hole mass 
$M_{{\rm tot}}$ in a double-peaked narrow emitter. For example, 
if $f_{{\rm 1}}=f_{{\rm 2}}=0.5$ (central two broad components have 
similar continuum emissions at 5100\AA) and 
$M_{{\rm 1}}\sim M_{{\rm 2}}$, we could have 
$M_{{\rm BH}}\sim M_{{\rm tot}}/2.8$. 

\begin{figure*}
\centering\includegraphics[width = 14cm,height=10cm]{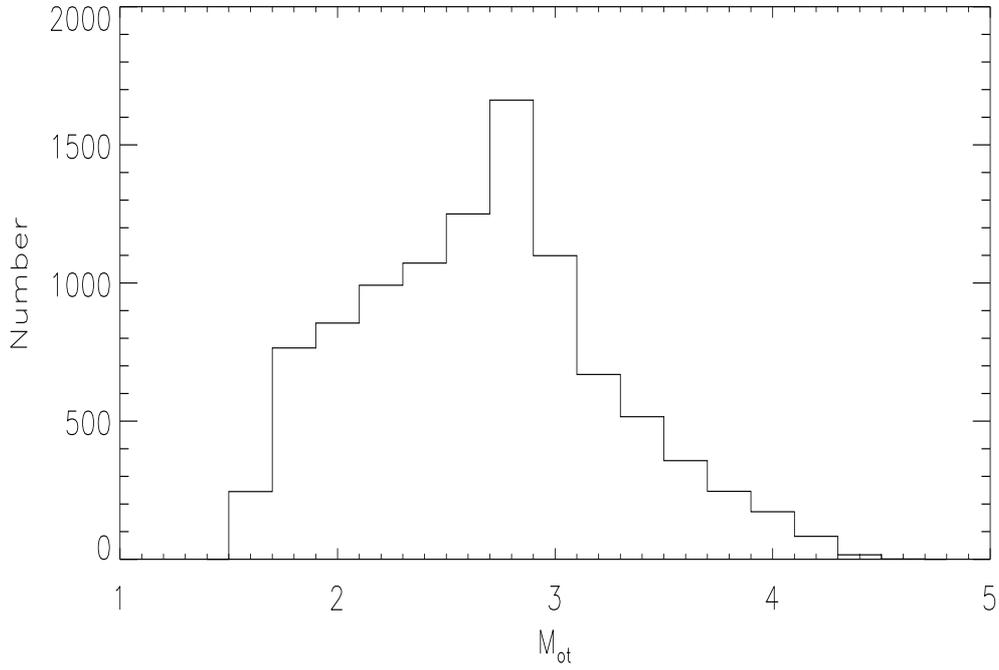}
\caption{On distribution of the black hole mass ratio $M_{{\rm ot}}$.
}
\label{dis_mock}
\end{figure*}

    Before the end of the Section, we give more clearer results 
on mass ratio ($M_{{\rm ot}}$) of $M_{{\rm tot}}$ to $M_{{\rm BH}}$. 
Based on ratios of two peak shifted velocities of double-peaked 
narrow emission lines of the reported double-peaked narrow 
emitters \citep*{wc09, ss10, gh12}, we 
simply accepted that $M_{{\rm 1}}/M_{{\rm 2}}$ is from 0.2 to 4 
under the DBH model. And, we accept that $f_{{\rm 1}}$ is 
from 0.2 to 0.8. Then, ten thousand values are randomly created for 
$M_{{\rm 1}}/M_{{\rm 2}}$, and for $f_{{\rm 1}}$. Distribution 
of the ten thousand calculated ratios of $M_{{\rm tot}}$ to 
$M_{{\rm BH}}$ is shown in Fig.~\ref{dis_mock}. It is clear that 
the mean value of $M_{{\rm ot}}$ is around 2.7 (minimum value 
larger than 1.5), and virial black hole masses from observed 
broad emission lines ($M_{{\rm BH}}$) should be statistically 
smaller than true virial black hole masses ($M_{{\rm tot}}$). 
Therefore, under the virialization assumption for central broad 
line regions and the DBH model for double-peaked narrow emission 
lines, it is interesting to check whether virial black hole 
masses of double-peaked narrow emitters are statistically 
smaller than normal broad line AGN with single-peaked narrow 
lines. 
 

\section{Main Results}

\subsection{Our Data Sample}

   Based on the samples of double-peaked narrow emitters reported 
in the literature, especially from the sample of \citet{ss10}, 
we can collect double-peaked narrow emitters with broad Balmer 
lines. Actually, in the large sample of 
\citet{gh12}, there are many emitters reported as type 1 AGN. However, 
when we checked their SDSS spectra, only weak broad components 
(second moment less than 800${\rm km/s}$) around H$\alpha$ can be 
found, no broad components around H$\beta$ can be found. It is hard 
to confirm the weak broad components from central BLRs of those 
objects. Thus, we mainly select targets from the double-peaked narrow 
emitters classified as QSOs in SDSS database \citep*{ya00, gs06, ew11, 
sg13, aa15}. Then, from low-redshift QSOs ($z<0.35$) in 
SDSS DR7 (Data Release 7, \citet{sr10}), 
there are 37 QSOs reported as double-peaked narrow emitters in 
\citet{ss10} included in our main sample. Here, the 
restriction of $z<0.35$ enables us to check both the broad 
H$\beta$ and the broad H$\alpha$, which will ensure the accuracy 
of following measured line widths of broad lines (sometimes 
effects of much extended wings of \oiii lines can not be clearly 
removed, if only broad H$\beta$ is checked). Basic information 
of 37 targets is listed in Table 1. 

   Then, as discussed in Section 2, it is necessary to determine 
line width of broad Balmer lines and continuum luminosity 
at 5100\AA, in order to estimate virial black hole masses. Here, 
one point we should note is that \citet{sr11} have reported 
virial black hole masses of QSOs in SDSS DR7 by continuum luminosity 
and FWHM of broad emission lines. In order to do convenient 
comparisons of virial black hole masses between the 37 double-peaked 
narrow emitters and normal broad line QSOs in \citet{sr11},  
FWHM is used as the line width of broad Balmer lines 
in this manuscript.

\begin{figure*}
\centering\includegraphics[width = 14cm,height=10cm]{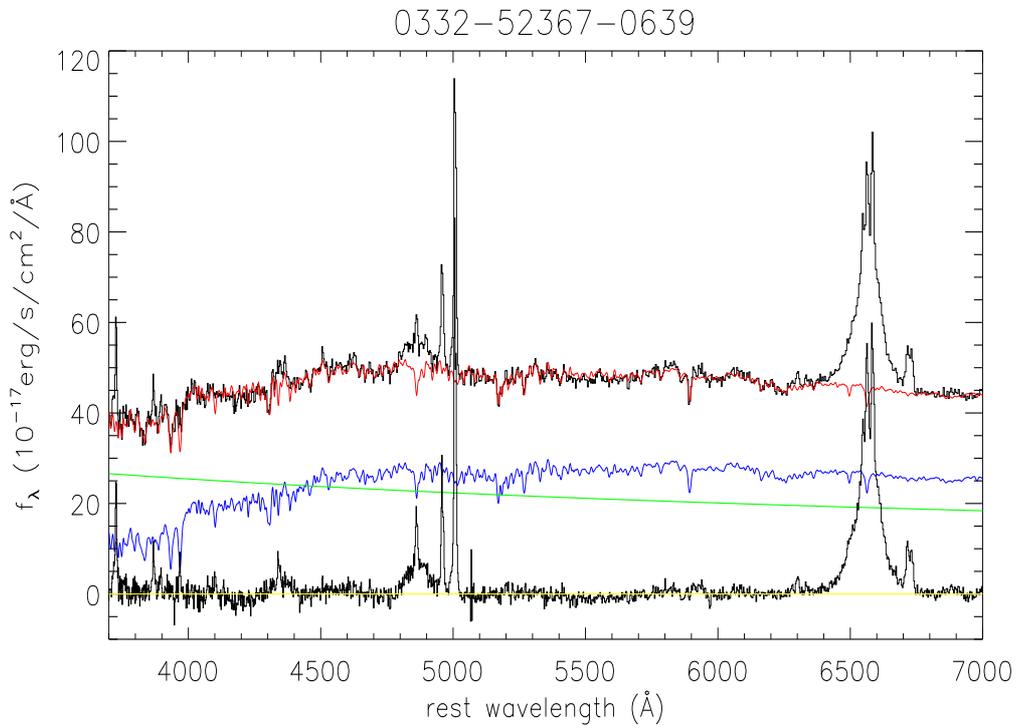}
\caption{An example shows the subtraction of stellar lights in SDSS 
0332-52367-0639 (plate-mjd-fiberid) by the SSP method. From top to bottom, 
solid lines in black and in red show the observed spectrum and the best 
fitted results respectively, solid lines in blue and in green show the 
determined stellar lights and the power law component respectively, solid 
lines in black and in yellow show the pure line spectrum and $f_\lambda=0$.
}
\label{ssp}
\end{figure*}

   Before measuring line parameters, we can find that there are 
9 of the 37 double-peaked narrow emitters, of which spectra include 
apparent contributions of star lights. Therefore, one procedure is 
first applied to subtract stellar lights in their spectra. Here, 
the SSP method (Simple Stellar Population) is applied with 39 
simple stellar template spectra from \citet{bc03} with stellar 
population ages from 5Myr to 12Gyr and with three metallicities 
(Z=0.008, 0.05, 0.02). More detailed and recent descriptions on 
the SSP method can be found in \citet*{bc03, cm05, rf09, cm12, zh14}  
etc.. Fig.~\ref{ssp} shows an example on the subtraction of stellar 
lights by the SSP method. Then, line parameters can be determined 
from line spectra after subtractions of stellar lights. 

    Emission lines around H$\alpha$ and H$\beta$ are then mainly 
focused on, and fitted simultaneously by the following 
model functions. Three (or more if necessary, after 
checking fitted results by three Gaussian functions) broad 
Gaussian functions are applied to describe each 
broad Balmer line, two Gaussian functions 
are applied to each double-peaked narrow emission line, and two 
additional Gaussian components are applied to probable 
extended wings of the \oiii doublet, one broad Gaussian  
function is applied to describe the weak He~{\sc ii} line, one 
power law function is applied to describe the AGN continuum 
emission, and then the Fe~{\sc ii} template discussed in 
\citet{kp10} is applied to describe the probable 
optical Fe~{\sc ii} lines. And moreover, if one narrow emission 
line is single-peaked, only one Gaussian function 
is applied to the narrow emission line. When the 
functions above are applied to fit the emission lines around 
H$\beta$ and H$\alpha$ simultaneously, the following restrictions 
are applied (parameters tied to one another in the MPFIT procedure), 
(1): blue (red) components of the double-peaked narrow emission 
lines have the same redshift, (2): corresponding 
broad components of broad H$\alpha$ and broad H$\beta$ have the 
same redshifts, when they are fitted by multiple broad 
Gaussian functions, (3): the flux ratios of components 
of the \oiii (the \nii) doublet are fixed to the theoretical values 
$f_{{\rm 5007}}/f_{{\rm 4959}}=3$ ($f_{{\rm 6585}}/f_{{\rm 6549}}=3$), 
(4): there are the same line widths of the blue (red) components of the 
double-peaked narrow Balmer lines (the \oiii or \oi or \sii 
doublets). Here, we should note that line flux of 
broad components are not tied between broad Balmer lines. In other 
words, different flux ratios are allowed for components of broad 
Balmer lines. And, in our procedure, not the severe restriction 
of 'the same line profile' but 'similar line profile' is applied 
to broad Balmer lines.

  Obviously, there are three main different points between our 
emission line fitting procedure and the procedure in \citet{sr11}. 
First and foremost, more than three Gaussian functions are allowed 
to fit broad Balmer lines, which can lead to more better fitted 
results for broad Balmer lines with more extended components. 
Besides, more recent and high-quality Fe~{\sc ii} template in 
\citet{kp10} is applied to describe optical band Fe~{\sc ii} 
components, rather than the template in \citet{bg92}. Last but 
not the least, the broad H$\alpha$ and H$\beta$ are fitted 
simultaneously, which can reduce effects of extended \oiii 
components on broad H$\beta$ as much as possible and lead to 
similar line profiles of broad Balmer lines (for example, by the 
procedure of \citet{sr11}, some QSOs have much different broad 
Balmer line widths as the results shown in following Fig.~\ref{par}).

\begin{figure*}
\centering\includegraphics[width = 18cm,height=18cm]{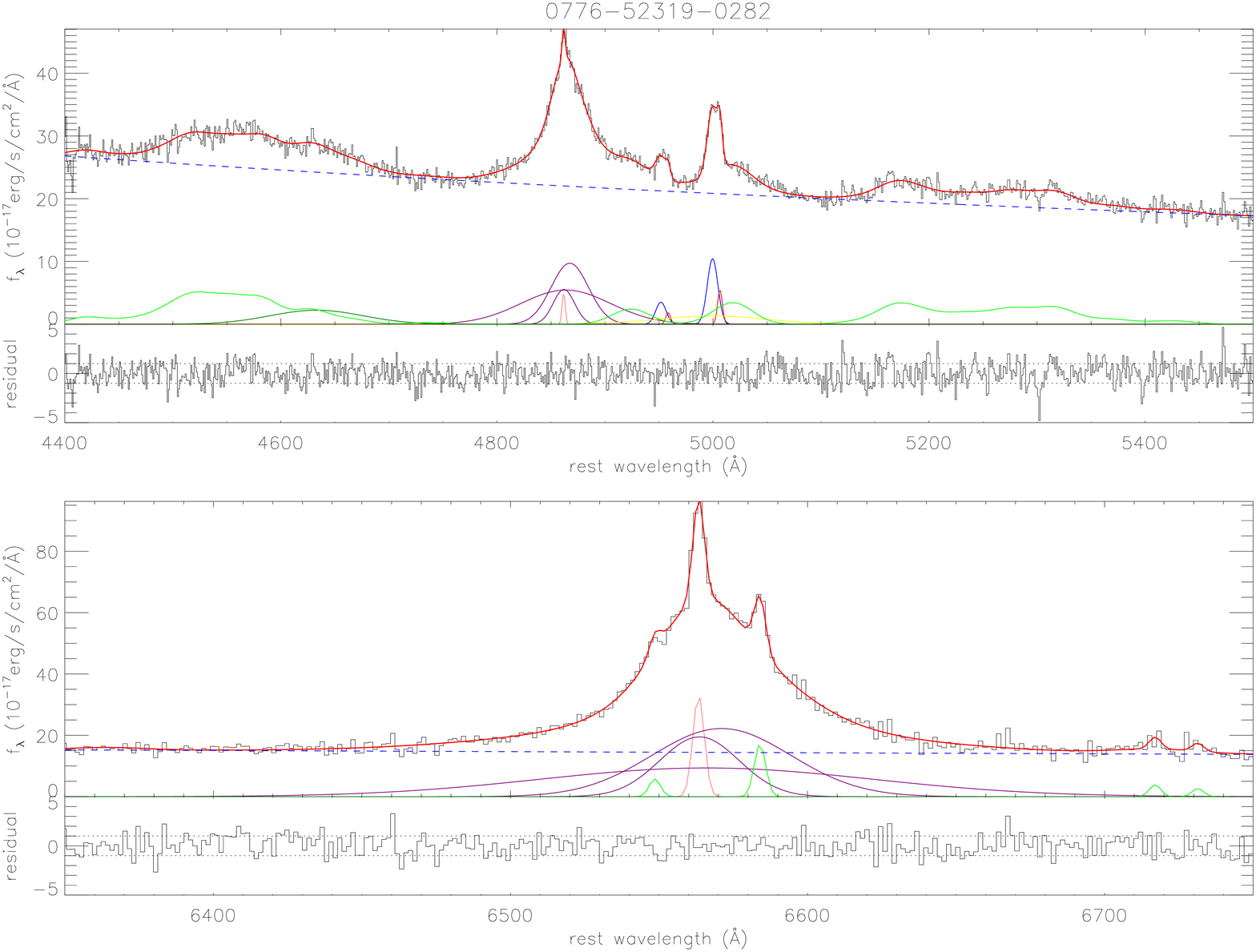}
\caption{Examples of the best fitted results for emission lines around 
H$\beta$ (top panel) and around H$\alpha$ (bottom panel) in SDSS 
0776-52319-0282. In each panel, top region shows the spectrum and the best 
fitted results, and bottom region shows the corresponding residuals. 
For the line spectrum and the best fitted results, solid line in black, 
thick solid line in red, solid lines in purple, in pink and dashed line 
in blue show the observed spectrum, the best fitted results, the 
determined broad Balmer components, the narrow Balmer component, 
and the AGN continuum emission 
respectively. And, for the best results for emission lines around 
H$\beta$, solid line in green, thin solid line in blue and in red, and 
thick solid line in dark green and in yellow show the Fe~{\sc ii} 
lines, the blue components and the red components of the \oiii 
doublet, the broad He~{\sc ii} line, and the extended components 
of the \oiii doublet respectively. And for the best fitted results 
for emission lines around H$\alpha$, solid line in green shows 
the \nii,\sii doublets. And for residuals, horizontal dotted lines 
show $residual=\pm1$.}
\label{line_ex}
\end{figure*}

    Then, through the Levenberg-Marquardt least-squares minimization 
technique (the MPFIT package, \citet{ma09}), the double-peaked 
and/or single-peaked narrow emission lines and the broad Balmer  
lines can be well determined. Here, we do not show the fitted results 
for emission lines of all the 37 double-peaked narrow emitters, but 
Fig.~\ref{line_ex} shows an example on the best fitted results for 
emission lines around H$\beta$ and around H$\alpha$ in SDSS 
0776-52319-0282 (plate-mjd-fiberid), of which there are double-peaked 
\oiii lines, but single-peaked other narrow emission lines, and apparent 
Fe~{\sc ii} lines. Then, for the 37 double-peaked narrow emitters, 
continuum luminosities at 5100\AA\ can be calculated by the determined 
power law AGN continuum emissions. And FWHMs of broad Balmer lines can 
be well determined by broad line profiles, after subtractions of the 
narrow emission lines, the power law continuum emissions,  
the Fe~{\sc ii} lines and the He~{\sc ii} lines.

    Then, we can check whether the determined broad line 
parameters are reliable. Here, we do not consider uncertainties 
of FWHM and continuum luminosity, because the uncertainties have few 
effects on statistic properties of virial black hole masses. 
The determined FWHMs, along with the determined line fluxes of 
broad Balmer lines and the continuum emission at 
5100\AA, are listed in Table 1. We first check line width correlation 
and line flux correlation between broad H$\beta$ and broad H$\alpha$, 
in order to ensure the accuracy of broad line parameters. 
Fig.~\ref{par} shows the correlations. It is clear that there are 
strong linear correlations between the broad Balmer  
lines. The Spearman rank correlation coefficients are about 0.96 
with $P_{{\rm null}}\sim10^{-21}$ and 0.88 with 
$P_{{\rm null}}\sim10^{-12}$ for the broad line width correlation 
and for the broad line flux correlation respectively. And moreover, 
the two correlations can be well described by 
$FWHM({\rm H\alpha})\sim 0.9\times FWHM({\rm H\beta})$ and  
$flux({\rm H\alpha})\sim3.3\times flux({\rm H\beta})$, which are  
consistent with the results shown in \citet{gh05}   
for QSOs. Therefore, the measured broad line widths and broad line 
fluxes are reliable. And moreover, in top panel of Fig.~\ref{par}, 
we also show broad line widths of 3477 normal broad line AGN from 
\citet{sr11}. We can find that there are more larger scatters 
for the broad line width correlation for the normal broad line AGN, 
without restriction of similar line profiles of broad Balmer  
lines. So that, the restriction of similar line profiles of broad 
Balmer lines can well ensure the accuracy of the 
measured broad line parameters.     
 
\subsection{Virial black hole masses}

    Based on the continuum luminosity at 5100\AA\  
($\lambda L_{{\rm 5100}}$) and the measured line with of broad 
H$\beta$, virial black hole masses of the 37 double-peaked narrow 
emitters can be estimated by the equation in \citet{vp06} (VP06) 
and in \citet{sr11}, 
\begin{equation}
\begin{split}
M_{{\rm BH}} & = A+B\log(\frac{\lambda L_{{\rm 5100}}}{10^{44}\ {\rm erg/s}}) +2\log(\frac{FWHM({\rm H\beta})}{\rm km/s})\\
& A = 0.91,\ \  B = 0.5
\end{split}
\end{equation}.
The main reason to use the equation in VP06 is mainly 
due to its application of 
$R_{{\rm BLR}}\propto(\lambda L_{{\rm 5100}})^{0.5}$ identical to 
more recent R-L relation in \citet{bd13}. The estimated virial 
black hole masses are also listed in the Table 1 for the 37 double-peaked 
narrow emitters. The mean value of virial black hole masses of the 37 
double-peaked narrow emitters is about 
$\log(M_{{\rm BH}}/{\rm M_{{\rm \odot}}})\sim8.11$.  

\begin{figure*}
\centering\includegraphics[width = 14cm,height=10cm]{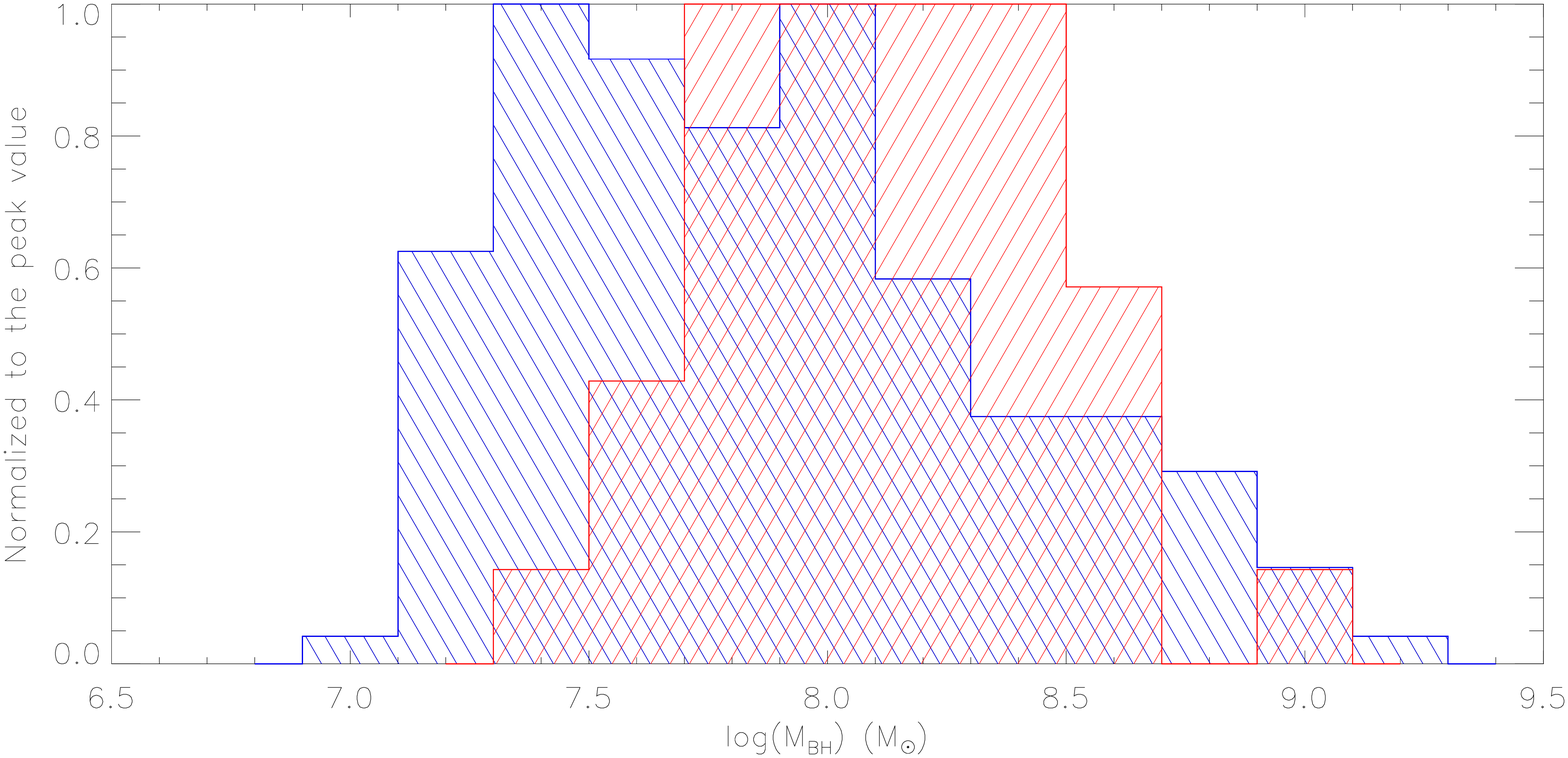}
\caption{Comparison of virial black hole masses between the 37 
double-peaked narrow emitters (in red color) and the 298 normal broad 
line AGN from SDSS DR7 (in blue color).
}
\label{com}
\end{figure*}

     Then, we can check whether are there different properties of 
virial black hole masses of normal broad line AGN with single-peaked 
narrow emission lines. A parent sample of normal broad line AGN can be 
created from the dataset of \citet{sr11}  by the the following 
three criteria, (1): objects are not double-peaked narrow emitters 
(information from key parameter of special\_interest\_flag), (2): objects 
have redshifts less than 0.35, (3): objects have similar measured 
broad Balmer line widths 
($0.8<\frac{FWHM({\rm H\alpha})}{FWHM({\rm H\beta})}=F_{ab}<1$, 
where range of [0.8, 1] is the range covered all the 37 double-peaked 
narrow emitters in the top panel of Fig.~\ref{par}), which will 
ensure the accuracy of broad Balmer line widths. Through the 
criteria (1) and (2), 3477 normal broad line AGN are selected, of 
which broad line widths of broad Balmer lines are 
shown in the top panel of Fig.~\ref{par}. And then, through the 
criterion (3), there are 298 normal broad line AGN included in our 
parent sample. The mean value is about 
$\log(M_{{\rm BH}}/{\rm M_{{\rm \odot}}})\sim7.88$ of the virial 
black hole masses estimated by the Equation (6) for the 298 normal 
broad line AGN in \citet{sr11}. And Fig.~\ref{com} shows 
distributions of virial black hole masses of the 37 double-peaked 
narrow emitters and the 298 normal broad line AGN. We can find 
clearly statistically larger virial black hole masses in the 37 
double-peaked narrow emitters with broad Balmer  
lines than in the normal broad line AGN.

   Then, the Student's T-statistic technique is applied 
to check whether the 37 double-peaked narrow emitters and the 
normal broad line AGN have significantly different mean virial 
black hole masses. For the parent sample including 298 normal broad 
line AGN and the sample of the 37 double-peaked narrow emitters, 
the calculated T-statistic value and its significance are 3.59 
and $7.1\times10^{-4}$ respectively, which indicate that the 37 
double-peaked narrow emitters and the normal broad line AGN have 
much different mean values of their black hole masses with confidence 
level higher than 99.9\%.  

\begin{figure*}
\centering\includegraphics[width = 14cm,height=10cm]{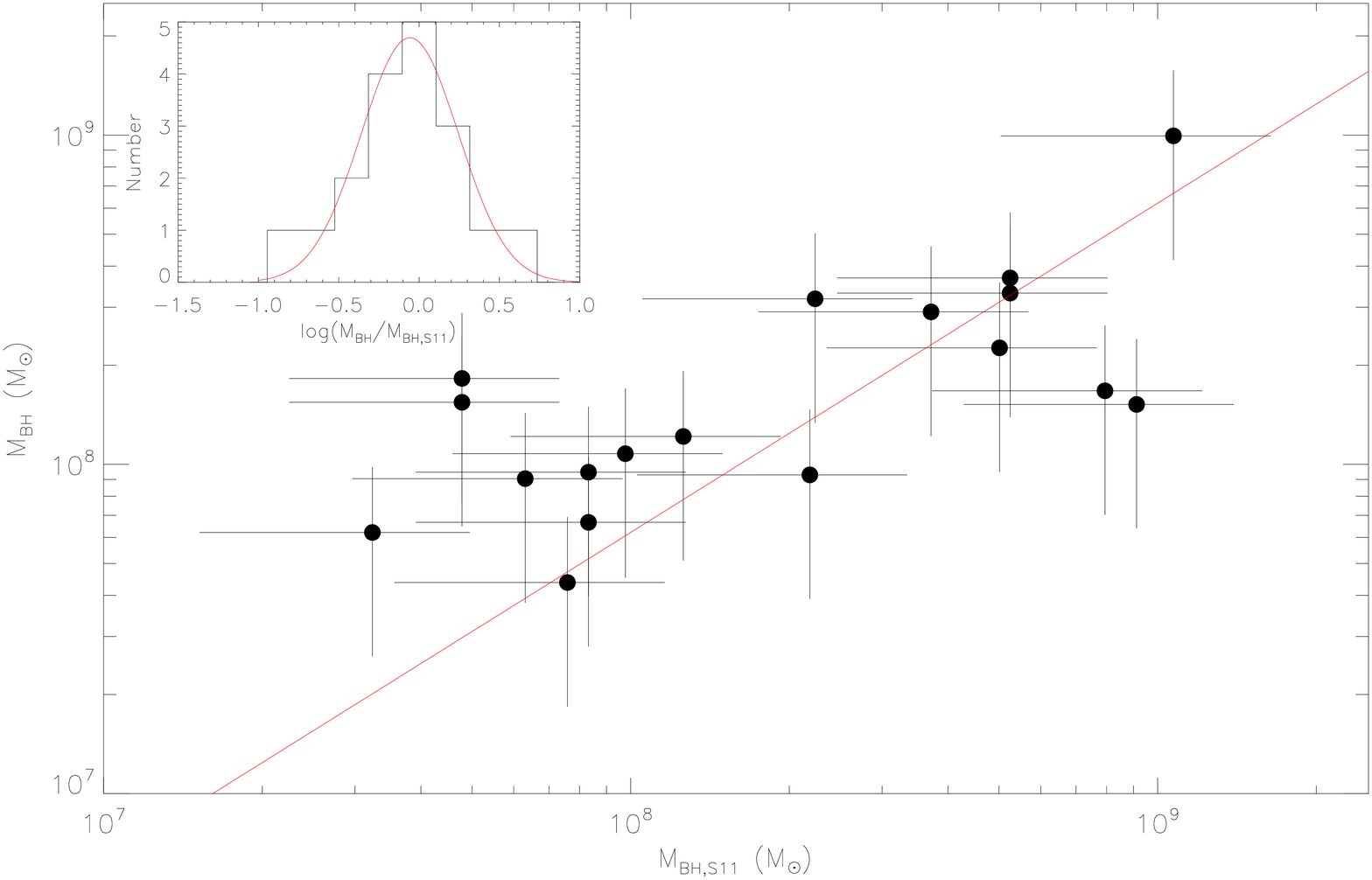}
\caption{Comparison of our determined virial black hole masses 
$M_{{\rm BH}}$ and the black hole masses $M_{{\rm BH,S11}}$ reported 
in Shen et al. (2011) for the 18 double-peaked narrow emitters 
included in Shen et al. (2011). The red solid line shows 
$M_{{\rm BH,S11}} = M_{{\rm BH}}$. Here, we assume the uncertainty of 
our determined virial black hole mass about 53\% (mean value from 
Shen et al. 2011). Top-left corner shows distribution of
$\log(M_{{\rm BH}}/M_{{\rm BH,S11}})$ which can be well described by
a Gaussian function with second moment 0.3 (solid line in red in the corner).
}
\label{mass2}
\end{figure*}

   Before proceeding further, we can find that among the 37 
double-peaked narrow emitters listed in Table 1, there are 18 
objects of which virial black hole masses can also be found in 
\citet{sr11}. Here, the 37 objects in our main 
sample are selected from 21592 QSOs with redshift less than 0.7\ 
in SDSS DR7 (see descriptions in \citet{ss10}), however, the 
catalogue of \citet{sr11} includes only 15798 QSOs with 
redshift less than 0.7. Therefore, not all the 37 objects are 
included in the QSO catalogue of \citet{sr11}. Although, the 
same equation in VP06 is applied to estimate the virial black hole 
masses of the 37 double-peaked narrow emitters and the normal 
broad line AGN in \citet{sr11}, a bit different procedures 
are applied to determine the line width of broad H$\beta$. Thus, it 
is necessary to check effects of different emission line fitting 
procedures on final virial black hole masses. Fig.~\ref{mass2} shows 
the comparison of our determined virial black hole masses 
$M_{{\rm BH}}$ and the reported masses $M_{{\rm BH,S11}}$ in 
\citet{sr11} for the 18 double-peaked narrow emitters. 
Here, values of $M_{{\rm BH,S11}}$ of the 18 objects are also 
listed in Table 1. The Spearman rank correlation coefficient for 
the correlation is about 0.67 with 
$P_{{\rm null}}\sim2\times10^{-3}$. And moreover, the 
top-left corner of Fig.~\ref{mass2} shows distribution of 
$\log(M_{{\rm BH}}/M_{{\rm BH,S11}})$ which can be well described 
by a Gaussian function with second moment 0.3. Therefore, 
$M_{{\rm BH}}\sim M_{{\rm BH,S11}}$ 
can be accepted for the 18 objects. Moreover, if we check the mass 
ratio of $M_{{\rm BH,S11}}$ to $M_{{\rm BH}}$, we will find the mean 
value of the ratio is about 1.62. Thus, we can safely accept that 
there are different mean virial black hole masses between the 37 
double-peaked narrow emitters and the 298 normal broad line AGN, even 
with considerations of effects of different emission line fitting 
procedures.

\begin{figure*}
\centering\includegraphics[width = 18cm,height=11cm]{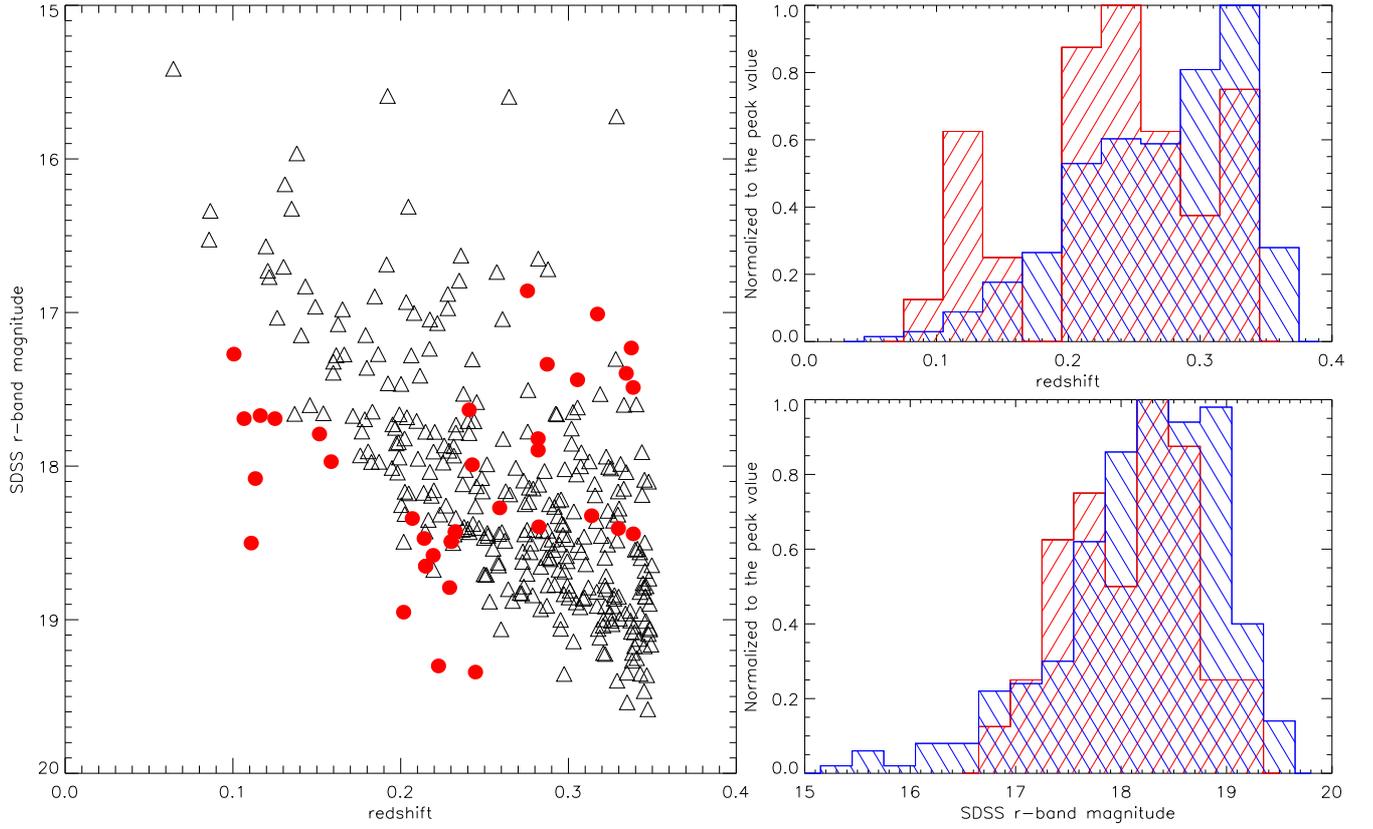}
\caption{Properties of redshift and SDSS r-band magnitude of the 37
double-peaked narrow emitters and the 298 normal broad line AGN.
Left panel shows the correlation between redshift and SDSS r-band
magnitude for the 37 double-peaked narrow emitters (red circles) and
the 298 normal broad emission line AGN (triangles). Top right panel 
shows the redshift distributions of the 37 double-peaked narrow 
emitters (red color) and the 298 normal broad line AGN (blue color). 
And bottom right panel shows the magnitude distribution of the 37 
double-peaked narrow emitters (red color) and the 298 normal broad 
line AGN (blue color).
}
\label{zmag}
\end{figure*}

     Moreover, we should note that there are different distributions 
of redshift and magnitude between the 37 double-peaked narrow emitters 
and the 298 normal broad line AGN. The distributions are shown in 
Fig.~\ref{zmag}. By two-sided Kolmogorov-Smirnov statistic technique, 
we can find that probability is less than 1\% that the 37 double-peaked 
narrow emitters and the 298 normal road line AGN have the same redshift 
distribution, and probability is less than 20\% that the 37 double-peaked
narrow emitters and the 298 normal road line AGN have the same SDSS r-band 
magnitude distribution.

\begin{figure*}
\centering\includegraphics[width = 18cm,height=14cm]{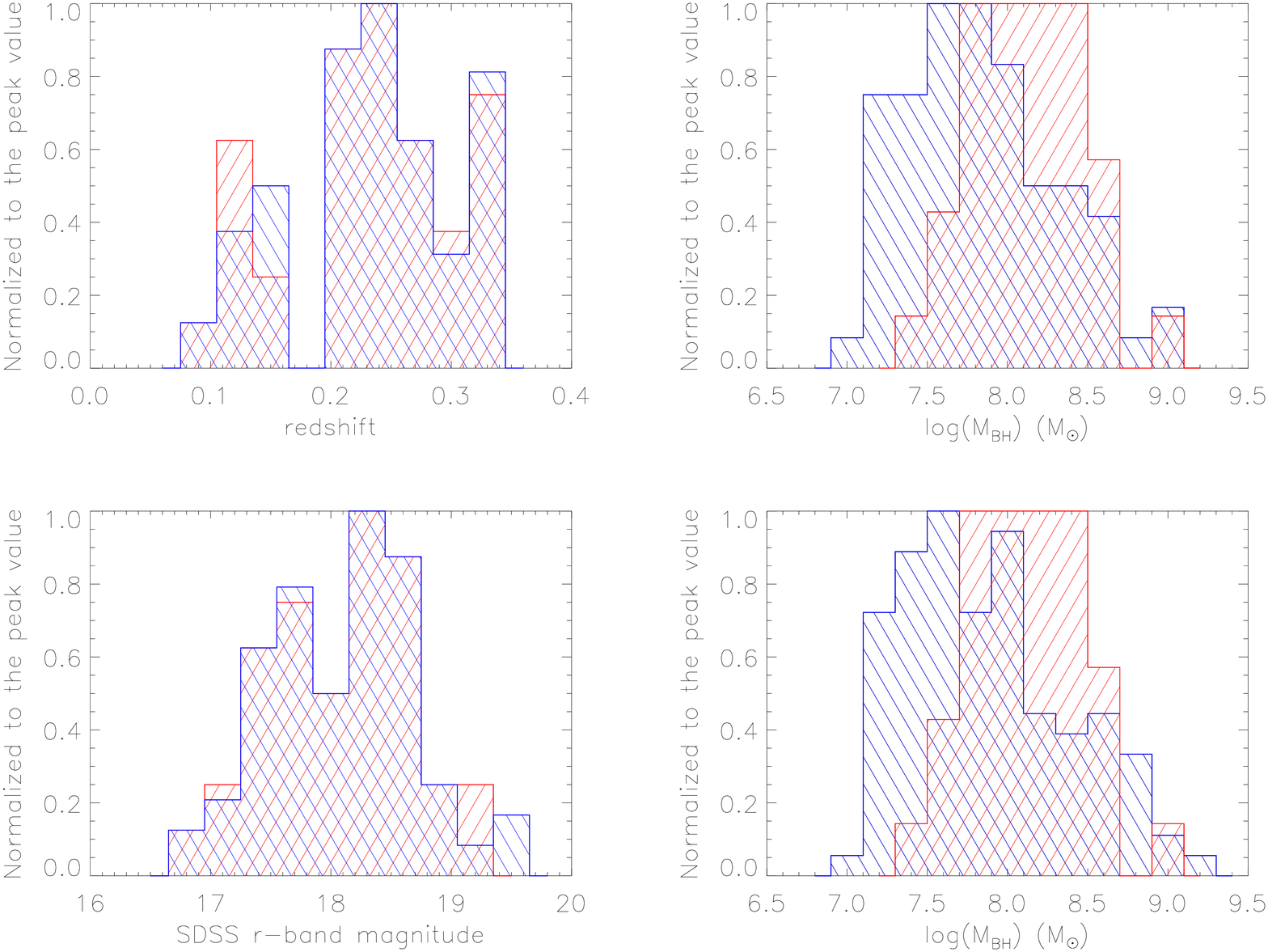}
\caption{Results on the first subsample and the second subsample. Top 
left and top right panels show the same redshift distributions and the 
comparison of black hole mass distributions of the 37 double-peaked 
narrow emitters and the first subsample of 74 normal broad line AGN 
respectively. Bottom left and bottom right panels show the same SDSS 
r-band magnitude distributions and the comparison of black hole mass 
distributions of the 37 double-peaked narrow emitters and the second 
subsample of 111 normal broad emission line AGN. In each panel, red and 
blue colors represent the results for the 37 double-peaked narrow emitters 
and for the objects in the subsamples respectively. 
}
\label{bh_dis2}
\end{figure*}

\begin{figure*}
\centering\includegraphics[width = 18cm,height=6cm]{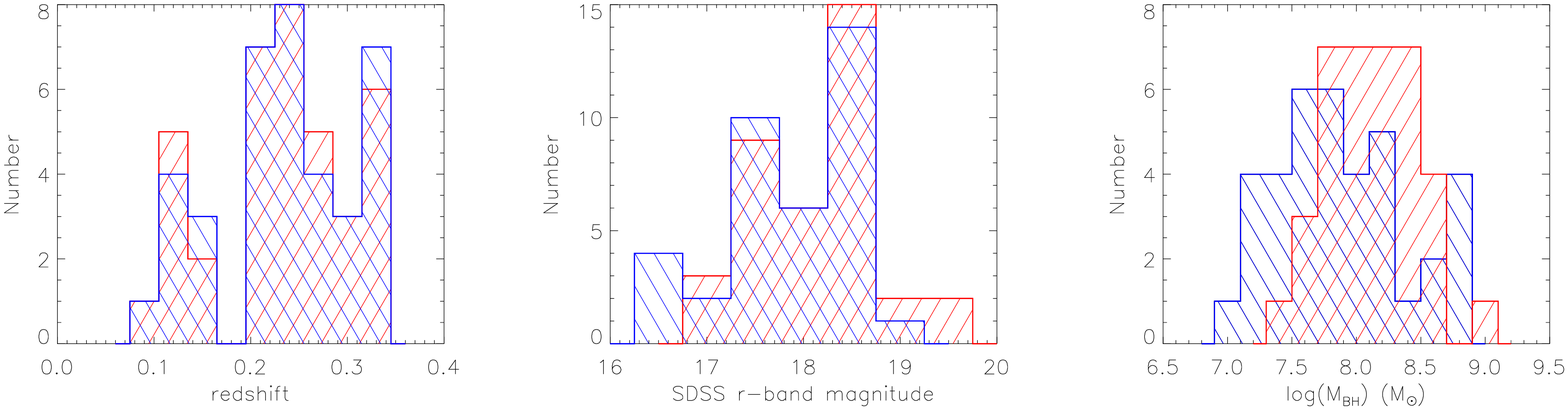}
\caption{Results on the third subsample with the same 
distributions of magnitude and redshift as those of the 37 double-peaked 
narrow emitters. Left, middle and right panels show the same redshift and 
the same magnitude distributions and the comparison of black hole mass 
distributions. In each panel, red and blue colors represent the results 
for the 37 double-peaked narrow emitters and for the objects in the 
third subsample respectively.
}
\label{bh_dis3}
\end{figure*}

    In order to consider effects of different distributions of redshift 
and/or magnitude on virial black hole mass comparisons as much as possible, 
three subsamples are created from the 298 normal broad line AGN as follows. 
To consider effects of different redshift distribution, the first 
subsample is created to include 74 normal broad line AGN, and the subsample 
has the same redshift distribution as that of the 37 double-peaked narrow 
emitters with probability larger than 92\%. The mean virial black hole 
mass of the subsample is about $\log(M_{{\rm BH}}/{\rm M_{\odot}})\sim7.84$. 
And the calculated T-statistic value and its significance are 3.39 and 
$1\times10^{-3}$ respectively for distributions of virial black hole masses 
of the 37 double-peaked narrow emitters and the normal broad line AGN in 
the subsample, which indicates that the 37 double-peaked narrow emitters 
and the normal broad line AGN in the subsample have much different mean 
virial black hole masses with confidence level higher than 99.8\%. In 
addition, in order to consider effects of different magnitude distribution, 
the second subsample is created to include 111 normal broad line AGN, 
and the subsample has the same SDSS r-band magnitude distribution  
as that of the 37 double-peaked narrow emitters with probability larger
than 99\%. The mean virial black hole mass of the subsample is about
$\log(M_{{\rm BH}}/{\rm M_{\odot}})\sim7.87$. And the the calculated T-statistic 
value and its significance are 3.23 and $2\times10^{-3}$ respectively 
for distributions of virial black hole masses of the 37 double-peaked 
narrow emitters and the normal broad line AGN in the subsample, which 
indicates that the 37 double-peaked narrow emitters and the normal 
broad line AGN in the subsample have much different mean virial black hole 
masses with confidence level higher than 99.7\%. The results on the 
first subsample and the second subsample are shown in Fig.~\ref{bh_dis2}. 
Furthermore, in order to consider effects of both different 
redshift distribution and different magnitude distribution, the third 
subsample is created to include 37 normal broad line AGN, and the 
subsample has both the same SDSS r-band magnitude distribution and the 
same redshift distribution as those of the 37 double-peaked narrow 
emitters with probability larger than 70\%. The mean virial black hole 
mass of the subsample is about $\log(M_{{\rm BH}}/{\rm M_{\odot}})\sim7.88$. 
And the the calculated T-statistic value and its significance are 
2.28 and $3\times10^{-2}$ respectively for distributions of virial 
black hole masses of the 37 double-peaked narrow emitters and the 
normal broad line AGN in the subsample, which indicates that the 37 
double-peaked narrow emitters and the normal broad line AGN in the 
subsample also have much different mean virial black hole masses with 
confidence level higher than 97\%. The results on the third subsample  
are shown in Fig.~\ref{bh_dis3}. Thus, with considerations of different 
redshift and/or magnitude distributions, the double-peaked narrow 
emitters have statistically larger virial black hole masses than the 
normal broad line AGN.

\subsection{Further discussions}

   In the subsection, there are three points we should note. First and 
foremost, the Equation (6) discussed in \citet{vp06}  is applied in 
the paper, due to its application of the R-L relation identical to 
the more recent results in \citet{bd13}. In order to confirm that 
there are few effects of different equations applied to estimate 
virial black hole masses on our final results. Further discussions 
are given as follows. Mean virial black hole mass of a sample of 
broad line AGN can be estimated by Equation (6) with different 
values of A and B 
\begin{equation}
\overline{M_{{\rm BH}}} = A + B\times\log(\overline{\frac{\lambda L_{{\rm 5100}}}{{10^{44}\ {\rm erg/s}}}}) + 2\times\log(\overline{FWHM})
\end{equation}.
Then, mean virial black hole mass of the 37 double-peaked narrow emitters 
is about $\overline{M_{{\rm BH, dbp}}}\sim A + 0.155\times B + 7.323$, 
and mean virial black hole mass is about 
$\overline{M_{{\rm BH, AGN}}}\sim A + 0.266\times B + 6.969$ for the 
normal broad emission line AGN in the third subsample discussed 
above with considerations of effects of different distributions of 
redshift and magnitude. In order to find smaller 
$\overline{M_{{\rm BH, dbp}}}$ than $\overline{M_{{\rm BH, AGN}}}$, we 
should have $B>3.2$. Similar results can be found with considering objects 
in the first subsample and the second subsample. However, more 
recent observational results in \citet{bd13} have shown 
that $B\sim0.533$. In other words, if we accepted the R-L relation 
$R_{{\rm BLR}}\propto \lambda L^{B}$ with $B\sim0.5$, it is hard to 
find smaller statistical virial black hole masses of the 37 
double-peaked narrow emitters.

    Besides, our main sample includes only 37 double-peaked narrow 
emitters, perhaps the larger virial black hole masses of the double-peaked 
narrow emitters are due to the selected double-peaked narrow emitters 
with larger broad line widths. The mean FWHM of broad H$\beta$ is about 
4580${\rm km/s}$ for the 37 double-peaked narrow emitters, however, the 
mean FWHM of broad H$\beta$ is 3000${\rm km/s}$ for the 298 normal 
broad emission line AGN in the parent sample. We try to discuss the 
different broad line widths as follows. On the one hand, when we select 
the 37 double-peaked narrow emitters, the objects with double-peaked 
\oiii lines classified as SDSS QSOs are firstly considered. Then, from 
the sample of double-peaked narrow emitters, the objects with broad 
Balmer lines are selected. Therefore, there are no effects 
of sample selection on the broad line width. On the other hand, under 
the DBH model for the double-peaked narrow emission lines, in order 
to explain the broader broad Balmer lines of the 37 double-peaked narrow 
emitters than normal broad line AGN (from 3000${\rm km/s}$ to 
4580${\rm km/s}$), the peak separation of central two broad components 
could be larger than 2000${\rm km/s}$, which is more larger than the 
reported peak separations of the double-peaked narrow emission lines. 
Therefore, although we have no clear ideas why there are broader 
broad Balmer lines in the double-peaked narrow emitters, 
the results on different broad line widths between the double-peaked 
narrow emitters and the normal broad emission line AGN can not support 
the DBH model. And more efforts should be done to enlarge the sample 
of the double-peaked narrow emitters with broad Balmer  
lines, in order to provide more further information on broad emission line 
properties.

    Last but not least, the main objective of this paper is to check 
the DBH model for the double-peaked narrow emitters, due to the expected 
smaller virial black hole masses. However, when we try to accept the 
smaller virial black hole masses in the double-peaked narrow emitters, 
there is one another assumption that the normal broad line AGN could 
not have common dual supermassive black holes with separation distance 
large enough to ensure two independent central BLRs. In other words, 
there is one unique BLR in central region of normal broad line AGN, 
however, there are two independent BLRs in central regions of the 
double-peaked narrow emitter under the DBH model. Therefore, even the 
BBH systems (binary black hole systems with separation distance about 
pcs or sub-pcs) are common in normal broad line AGN, smaller virial 
black hole masses through the broad Balmer line width and the continuum 
luminosity could be expected for the double-peaked narrow emitters. 
 
\section{Conclusions}

    Finally, we give our main conclusions as follows. On the one hand, 
through the kinematic model on the dual supermassive black holes, smaller 
central virial black hole masses could be expected in the double-peaked 
narrow emitters than in the normal broad line AGN, through the 
virialization method applied with the observed broad Balmer line width 
and the observed continuum luminosity. On the other hand, the virial 
black hole mass comparisons between the double-peaked narrow emitters 
and the normal broad line AGN show statistically larger virial black 
hole masses in the double-peaked narrow emitters, which are against 
the expected results by the DBH model. Therefore, the model on the dual 
supermassive black holes is not statistically preferred to the 
double-peaked narrow emitters.  

\section*{Acknowledgements}
Zhang and Feng gratefully acknowledge the anonymous referee for 
giving us constructive comments and suggestions to greatly improve our paper.   
Zhang acknowledges the kind support from the Chinese grant NSFC-U1431229. 
FLL is supported under the NSFC grants 11273060, 91230115 and 11333008, 
and State Key Development Program for Basic Research of China 
(No. 2013CB834900 and 2015CB857000). This paper has made use of the 
data from the SDSS projects. Funding for SDSS-III has been provided by 
the Alfred P. Sloan Foundation, the Participating Institutions, the 
National Science Foundation, and the U.S. Department of Energy Office 
of Science. The SDSS-III web site is http://www.sdss3.org/. SDSS-III 
is managed by the Astrophysical Research Consortium for the Participating 
Institutions of the SDSS-III Collaboration including the University of 
Arizona, the Brazilian Participation Group, Brookhaven National Laboratory, 
Carnegie Mellon University, University of Florida, the French 
Participation Group, the German Participation Group, Harvard University, 
the Instituto de Astrofisica de Canarias, the Michigan State/Notre 
Dame/JINA Participation Group, Johns Hopkins University, Lawrence 
Berkeley National Laboratory, Max Planck Institute for Astrophysics, 
Max Planck Institute for Extraterrestrial Physics, New Mexico State 
University, New York University, Ohio State University, Pennsylvania 
State University, University of Portsmouth, Princeton University, the 
Spanish Participation Group, University of Tokyo, University of Utah, 
Vanderbilt University, University of Virginia, University of Washington, 
and Yale University.

\begin{table*}
\begin{minipage}{16cm}
\caption{Parameters of the 37 double-peaked narrow emitters}
\begin{tabular}{lcccllclcc}
\hline
pmf & z & Mag & $\lambda L$ & $V_\beta$ & $f_\beta$ &
  $V_\alpha$ & $f_\alpha$ & $\log(M_{{\rm BH}})$ & $\log(M_{{\rm BH,S11}})$ \\
\hline
0332-52367-0639*   &   0.100   &   17.27   &   43.44   &   5510   &   750$\pm$111   &   4512   &   3712$\pm$331   &   8.11  & \\
0452-51911-0080   &   0.158   &   17.97   &   43.21   &   7029   &   308$\pm$86   &   7034   &   1971$\pm$502   &   8.21   & \\
0472-51955-0101   &   0.222   &   19.30   &   43.20   &   8938   &   278$\pm$57   &   6401   &   1512$\pm$151   &   8.41   &\\
0512-51992-0632   &   0.240   &   17.63   &   44.27   &   5446   &   2238$\pm$92   &   5162   &   8247$\pm$236   &   8.52   & 8.72\\
0553-51999-0048*   &   0.219   &   18.58   &   43.16   &   3308   &   376$\pm$55   &   3389   &   1691$\pm$131   &   7.53   & \\
0605-52353-0126   &   0.329   &   18.40   &   44.21   &   2046   &   501$\pm$35   &   1948   &   1594$\pm$111   &   7.64  & 7.88 \\
0607-52368-0625   &   0.275   &   16.85   &   44.71   &   3134   &   2270$\pm$113   &   3373   &   6478$\pm$267   &   8.26  & 7.68  \\
0609-52339-0435   &   0.230   &   18.49   &   43.95   &   3493   &   519$\pm$57   &   3786   &   1268$\pm$129   &   7.97   & 7.92\\
0616-52374-0415   &   0.214   &   18.65   &   43.70   &   6262   &   237$\pm$72   &   6005   &   1231$\pm$309   &   8.35   & \\
0616-52442-0437   &   0.214   &   18.65   &   43.78   &   6540   &   425$\pm$32   &   5654   &   1413$\pm$77   &   8.43   & \\
0623-52051-0224   &   0.116   &   17.67   &   43.72   &   5324   &   1860$\pm$339   &   3830   &   5717$\pm$1228   &   8.22 &  \\
0776-52319-0206   &   0.232   &   18.42   &   43.55   &   3706   &   340$\pm$57   &   3584   &   1889$\pm$728   &   7.82   & 7.92\\
0776-52319-0282   &   0.338   &   17.48   &   44.59   &   2589   &   1126$\pm$167   &   2257   &   3268$\pm$539   &   8.03   & 7.99\\
0793-52370-0293   &   0.305   &   17.43   &   44.53   &   4950   &   1974$\pm$549   &   4099   &   6322$\pm$1564   &   8.56   & 8.72\\
0978-52431-0633   &   0.281   &   17.82   &   44.30   &   2851   &   703$\pm$83   &   2675   &   2713$\pm$553   &   7.97   & \\
0978-52441-0622   &   0.281   &   17.89   &   44.31   &   2310   &   843$\pm$70   &   2329   &   2364$\pm$224   &   7.79   & 7.51\\
1295-52934-0580   &   0.242   &   17.99   &   44.19   &   5578   &   1118$\pm$94   &   4677   &   2708$\pm$210   &   8.50   & 8.35\\
1426-52993-0110   &   0.282   &   18.39   &   44.09   &   4285   &   794$\pm$85   &   4223   &   3542$\pm$285   &   8.22   & 8.90\\
1446-53080-0266   &   0.317   &   17.01   &   44.68   &   3553   &   3399$\pm$269   &   3949   &   12593$\pm$863   &   8.35   &8.70\\
1605-53062-0443   &   0.313   &   18.32   &   44.23   &   2914   &   565$\pm$95   &   2400   &   2142$\pm$334   &   7.95   & 7.80\\
1622-53385-0533*   &   0.201   &   18.95   &   43.10   &   4219   &   72$\pm$36   &   3710   &   572$\pm$362   &   7.71   & \\
1643-53143-0532   &   0.259   &   18.27   &   44.19   &   3464   &   969$\pm$162   &   2919   &   2660$\pm$529   &   8.08   & 8.10\\
1716-53827-0140*   &   0.151   &   17.79   &   43.56   &   8071   &   522$\pm$112   &   6156   &   1870$\pm$275   &   8.50   & \\
1762-53415-0388*   &   0.113   &   18.08   &   43.15   &   4742   &   368$\pm$84   &   4381   &   3295$\pm$503   &   7.84  & \\
1762-53415-0597   &   0.287   &   17.33   &   44.48   &   3297   &   1436$\pm$130   &   3745   &   3676$\pm$315   &   8.18  & 7.68 \\
1853-53566-0561   &   0.334   &   17.39   &   44.55   &   8013   &   1827$\pm$92   &   7251   &   7731$\pm$432   &   8.99   & 9.03\\
2001-53493-0154   &   0.214   &   18.47   &   43.94   &   5456   &   526$\pm$92   &   5456   &   1271$\pm$222   &   8.35  & \\
2004-53737-0634   &   0.125   &   17.69   &   43.79   &   3892   &   1534$\pm$75   &   3077   &   3945$\pm$148   &   7.99  & \\
2022-53827-0553*   &   0.206   &   18.34   &   42.99   &   7732   &   517$\pm$205   &   7804   &   2144$\pm$830   &   8.18  & 8.96 \\
2341-53738-0523   &   0.338   &   18.44   &   44.27   &   5103   &   668$\pm$46   &   4455   &   2179$\pm$151   &   8.46  & 8.57 \\
2365-53739-0359*   &   0.110   &   18.50   &   43.23   &   4325   &   815$\pm$89   &   4313   &   3224$\pm$340   &   7.79   & \\
2527-54569-0262   &   0.229   &   18.79   &   43.89   &   6386   &   430$\pm$58   &   5366   &   1371$\pm$132   &   8.46   & \\
2776-54554-0251*   &   0.106   &   17.69   &   43.29   &   3520   &   937$\pm$104   &   3211   &   5058$\pm$329   &   7.64  & \\
2793-54271-0614*   &   0.244   &   19.34   &   43.31   &   2356   &   342$\pm$34   &   2342   &   1116$\pm$146   &   7.31   &\\
2947-54533-0050   &   0.230   &   18.49   &   43.81   &   2794   &   460$\pm$51   &   1960   &   1132$\pm$156   &   7.70   &\\
2953-54560-0530   &   0.337   &   17.23   &   44.70   &   2257   &   1169$\pm$168   &   1900   &   4659$\pm$626   &   7.96  & 8.34 \\
3211-54852-0404   &   0.232   &   18.43   &   43.34   &   4325   &   153$\pm$9   &   3959   &   701$\pm$104   &   7.85   & \\
\hline
\end{tabular}\\
Note: The first column shows the SDSS plate-mjd-fiberid, 
the second column shows the redshift, the third column shows the 
SDSS r-band psf magnitude, the fourth column shows the continuum 
luminosity at 5100\AA\ ($\log(\lambda L_{{\rm 5100}})$) in unit of 
${\rm erg/s}$, the fifth and the sixth columns show the line width 
in unit of ${\rm km/s}$ and the line flux in unit of 
${\rm 10^{-17}erg/s/cm^2}$ of the broad H$\beta$, the seventh and the 
eighth columns show the line width and the line flux of the broad 
H$\alpha$, the ninth column shows the virial black hole mass 
$\log(M_{{\rm BH}})$ in unit of ${\rm M_\odot}$, the final column shows 
the virial black hole mass $\log(M_{{\rm BH,S11}})$ in unit of 
${\rm M_\odot}$ reported in Shen et al. (2011). \\
Symbol of * means the SDSS spectrum of the object includes apparent 
contributions of stellar lights.
\end{minipage}
\end{table*}
\clearpage

\begin{figure*}
\centering\includegraphics[width = 12cm,height=16cm]{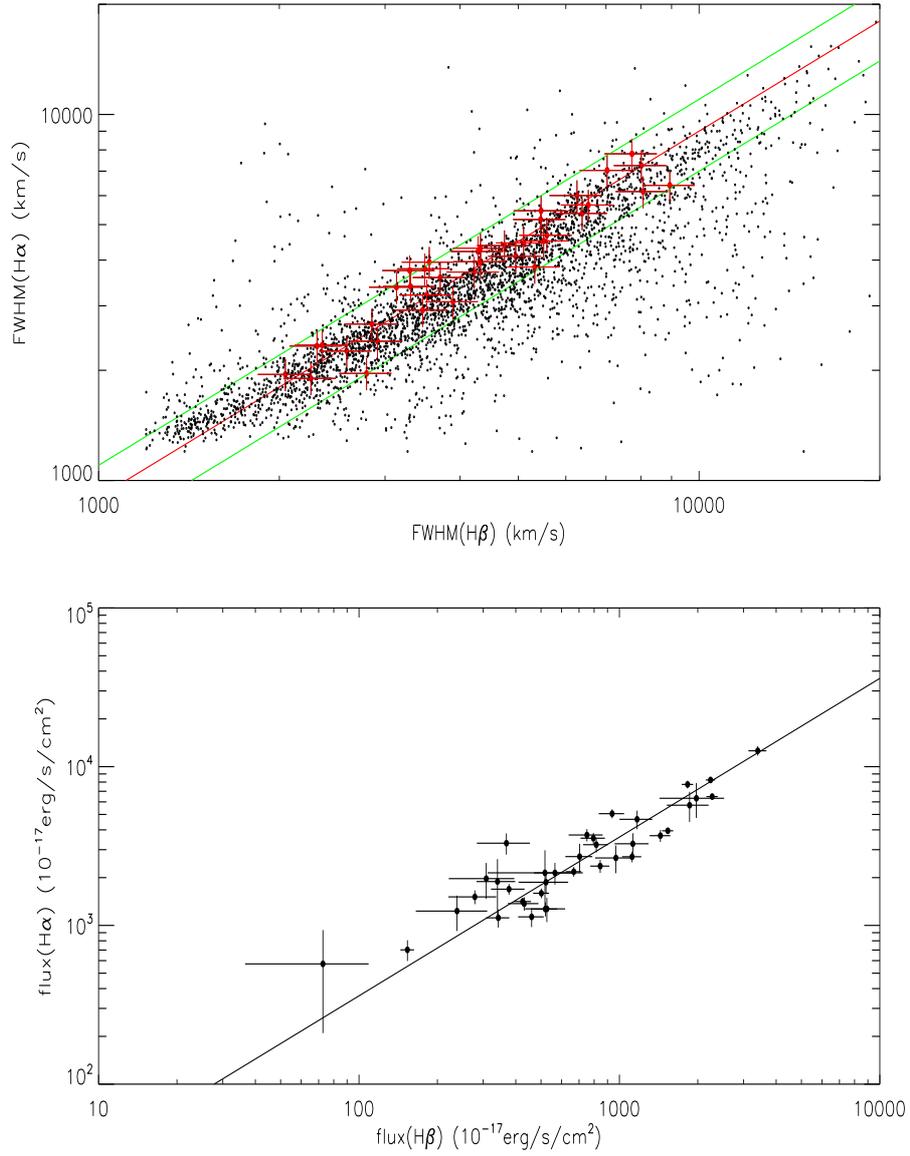}
\caption{On broad line width correlation and broad line flux 
correlation between broad H$\beta$ and broad H$\alpha$. In top panel, 
red circles are for the 37 double-peaked narrow emitters, dots are 
for the 3477 normal broad line AGN from SDSS DR7, and solid line 
in red shows $FWHM({\rm H\alpha})=0.91\times FWHM({\rm H\beta})$. And the two 
green lines represent $FWHM({\rm H\beta})=0.8\times FWHM({\rm H\alpha})$ and 
$FWHM({\rm H\beta})=1\times FWHM({\rm H\alpha})$ respectively. In bottom panel, 
solid line shows $flux({\rm H\alpha}) = 3.3\times flux({\rm H\beta})$.
}
\label{par}
\end{figure*}

\label{lastpage}
\end{document}